# Quality Analysis of Battery Degradation Models with Real Battery Aging Experiment Data

| Cunzhi Zhao | Xingpeng Li | Yan Yao |
|---|---|---|
| *Student Member, IEEE* | *Senior Member, IEEE* | *Senior Member, IEEE* |
| University of Houston | University of Houston | University of Houston |
| Houston, TX, USA | Houston, TX, USA | Houston, TX, USA |
| czhao20@uh.edu | xli82@uh.edu | yyao4@uh.edu |

***Abstract*—The installation capacity of energy storage system, especially the battery energy storage system (BESS), has increased significantly in recent years, which is mainly applied to mitigate the fluctuation caused by renewable energy sources (RES) due to the fast response and high round-trip energy efficiency of BESS. The main components of majority of BESSs are lithium-ion batteries, which will degrade during the BESS daily operation. Heuristic battery degradation models are proposed to consider the battery degradation in the operations of energy systems to optimize the scheduling. However, those heuristic models are not evaluated or demonstrated with real battery degradation data. Thus, this paper will perform a quality analysis on the popular heuristic battery degradation models using the real battery aging experiment data to evaluate their performance. A benchmark model is also proposed to represent the real battery degradation value based on the averaged cycle value of the experimental data.**

## I. Introduction

The decarbonization trend leads to the new challenge in power systems, which is the increased uncertainty associated with the large amount of renewable energy sources deployed in the system [1]. Thus, battery energy storage system (BESS) develops fast in recent years to address the uncertainty and inefficiency issues that caused by renewable energy sources (RESs) due to its fast response and high round-trip efficiency [2]. A good example is that there are a number of new solar and battery co-located farms recently [3].

Previous studies have proved that BESS can be a perfect solution to deal with the uncertainty caused by RESs [4]-[7]. However, none of those papers consider the battery degradation of the BESS in their energy management strategy. The main component of the majority types of BESS in the current market is lithium-ion battery cell. Lithium-ion batteries are connected in parallel and/or series in the battery modules. A BESS battery pack consists of multiple battery modules. The chemical characteristics of lithium-ion batteries make it degrade during the cycling [8]. This could lead to huge battery degradation over the years and result to financial losses for investors if it is not considered in the energy management system.

Thus, some papers proposed heuristic battery degradation models (BDMs) to mitigate the gap that the battery degradation can be considered in the energy management system. References [9]-[11] proposed a linear degradation cost parameter that is related to the power or energy usage. In other words, they added a linear battery degradation cost in the objective function in the scheduling optimization problem. A battery degradation model based on the depth of discharge (DOD) of each cycle is proposed in [12]-[14]. The degradation is calculated based on the average degradation value of each cycle respect to the experimental data that under certain DOD. These heuristic models seems reasonable and effective in the battery degradation quantification. However, they are not evaluated and compared with the real battery aging experiment data.

To evaluate the performance of the popular heuristic BDMs, this paper conducted a quantity analysis of those BDMs with the real battery aging experiment data. A benchmark model is also created based on the true degradation from the experiment data to gauge the accuracy of the heuristic BDMs. Also, the real data analysis is conducted in this paper. Therefore, we will be able to verify the performance of the heuristic BDMs and learn the degradation characteristic of the lithium-ion batteries.

The remainder of the paper is organized as follows. The mathematical formulations for heuristic popular BDMs are presented in Section II. Section III describes the details and analysis of the real battery aging experiment data. Model comparison and result discussions are presented in Section IV. Section V concludes the paper.

## II. Heuristic Battery Degradation Models

In this paper, two popular BDMs are selected to perform a quality analysis with the real battery degradation data. The formulation of BDMs are shown below in this section.

### *A. Linear Degradation Model*

The linear degradation model with a constant degradation rate [9]-[11]. The battery degradation cost (BDC) in their proposed BDM is either linear with the power usage or the energy usage of BESS. The constant degradation rate is determined with the manufacture battery data: dividing the capital cost by the projected total available energy as shown in (1), where the $Cycle$ represents the predicted lifecycle number, $E_{BESS}$ represents the maximum energy capacity of BESS and $c_{BESS}$ represents the total investment costs of BESS. The battery degradation cost is shown in (2) where $P_{BESS}^{Charge,t}$ and $P_{BESS}^{Discharge,t}$ represent the charging and discharging power at time period t respectively.

$$c_{BD} = c_{BESS}/(Cycle * E_{BESS}) \tag{1}$$

$$f(BDC) = \sum_t c_{BD} * (P_{BESS}^{Charge,t} + P_{BESS}^{Discharge,t}) \qquad (2)$$

*B. DOD based Degradation Model*

The second BDM that proposed in [12]-[14] is based on the DOD value for each cycle. The degradation value for each cycle is determined by the average cycle degradation under that certain DOD at the experiment data. The DOD value versus predicted number of lifecycle is shown in Fig. 1 [15]. Different DOD values will lead to the different numbers of cycles to reach a certain percent of the original capacity. In other words, the degradation value will be different under different DOD values. The $c_{BD}^{DOD,t}$ shown in (3) in this BDM represents the degradation cost variable that is determined by different DOD value based on the battery degradation data.

$$f(BDC) = \sum_t c_{BD}^{DOD,t} \qquad (3)$$

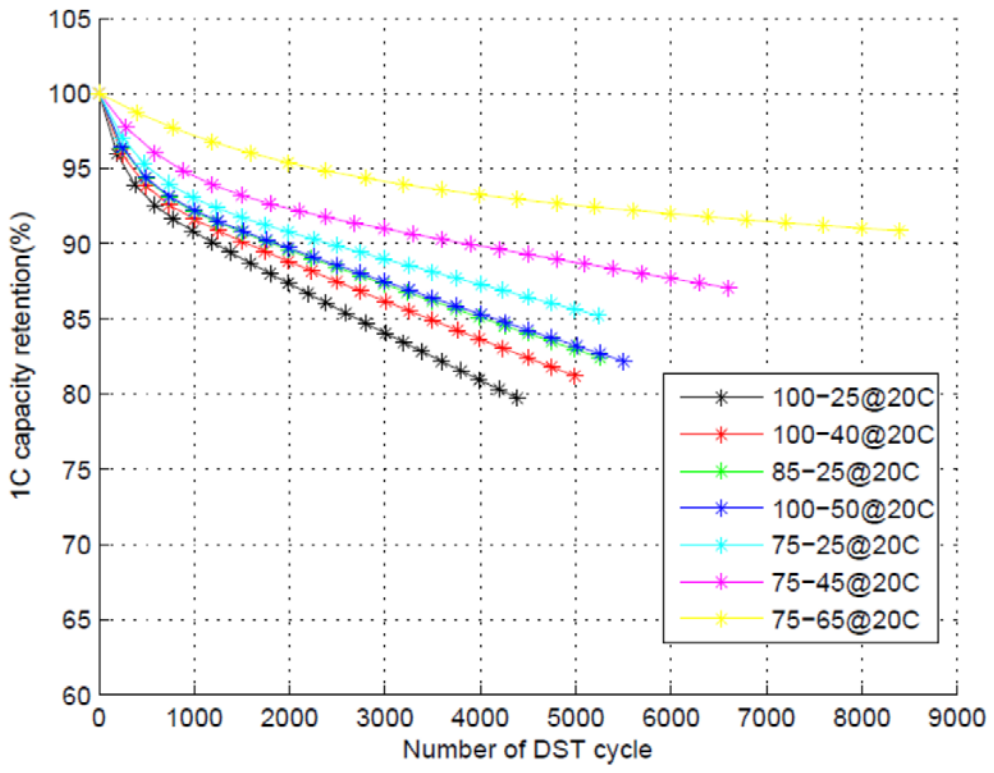

Figure 1. Dynamic stress test (DST) under different DOD values [15].

## III. Battery Degradation Data

The real battery degradation dataset analyzed in this paper is obtained from the Battery Archive website [16] that is maintained by Sandia National Lab. It has data for over 100 lithium-ion battery aging tests, which are publically available and open source for use. Those battery aging tests covers different ambient temperature, charge or discharge rate, DOD values and different materials of cathode for the lithium-ion batteries. Table I presents the numbers of the battery aging tests under different discharge rates.

Table I Distribution of battery aging tests.

| Battery Aging Tests # | Discharge Rate | | | | |
|---|---|---|---|---|---|
| Cathode Type | 0.5C | 1C | 1.5 C | 2C | 3C |
| LCO | 7 | / | / | 8 | / |
| NMC-LCO | / | / | 15 | / | / |
| LFP | 7 | 9 | / | 6 | 7 |
| NCA | 29 | 10 | / | 6 | / |
| NMC | 8 | 10 | / | 6 | 8 |

Fig. 2 represents a group of battery aging tests that were tested under 100% DOD, and 1.5C discharge rate at 25℃ ambient temperature. The results of six independent and identical tests that are under same testing condition are presented in Fig. 2. This figure gives an overview of the battery degradation results. The *y* axis represents the capacity of the battery and the *x* axis represents the cycle number. The capacity is determined by the available fully discharge capacity at each cycle. The degradation is calculated by the difference in energy capacities between two continuous cycles. From the figure, even though the six tests operates at the same condition, we can observe that the capacity curve does not overlap. Each battery performs different especially after the 700 cycles, this is due to the stochastic characteristic of lithium-ion battery [17]. Also, there are a lot of spikes from the capacity curve, those spikes are resulted by procedures of battery aging tests. After a certain number of cycling, they conduct a lower discharge rate for 1 cycle which results to a higher available discharge capacity for that certain cycle. In this case, the normal discharge rate is 1.5C while it is 0.5C for those spikes' cycle. Also, there are some cycles' capacities dropping down to 0, which is because those cycles are in the idle mode. Thus, when we analyze the tests, we can ignore those spikes in the following figure. Fig. 3 shows the detail of Fig. 2 by limiting the *y* axis between 0.5 and 3 Ah. From Fig. 3, it is more clearly to observe the capacity degrades diversely even under the same aging test condition with the same battery. This also indicates the difficulty of the battery degradation prediction.

Similar to Figs. 2 and 3, we also present another groups of battery aging tests with lithium ferro-phosphate battery (LFP). Fig. 4 shows the battery aging data of 4 groups of tests. From the figure, it seems the LFP is even worse in terms of performance consistency and stability. Only tests *c* and *d* follow the expected trend of the expected capacity curve. It is extreme unstable for tests *a* and *b*. Thus, we can conclude that different types of lithium-ion batteries may perform very differently.

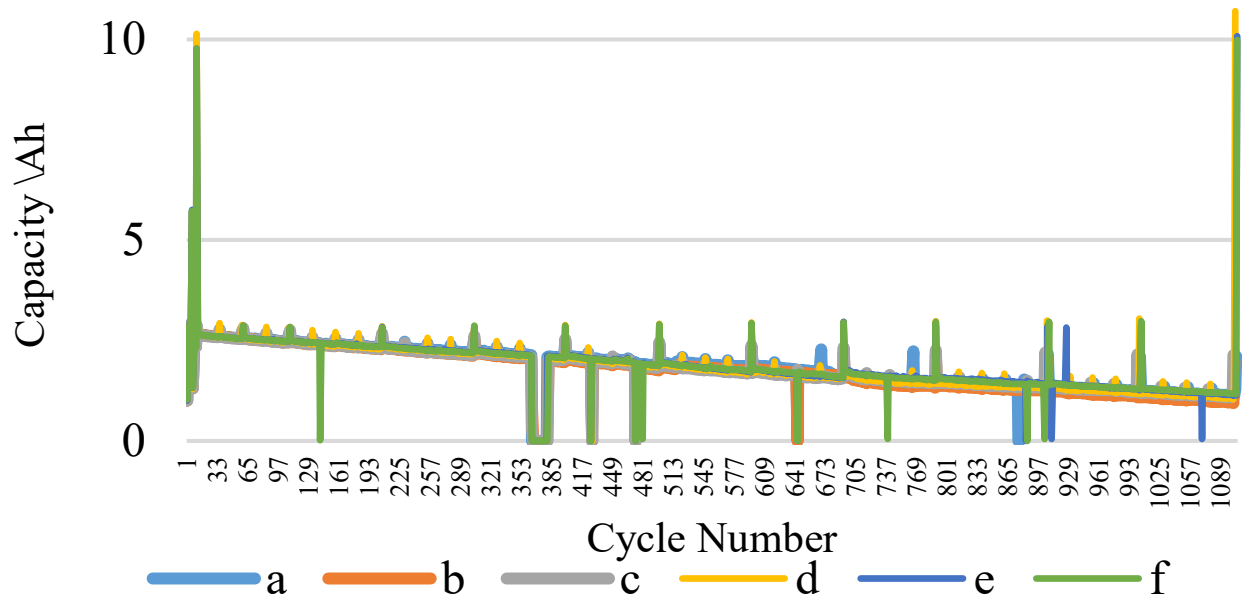

Figure 2. Battery capacity curve of HNEI_18650_NMC_LCO_25C_0-100_0.5-1.5C.

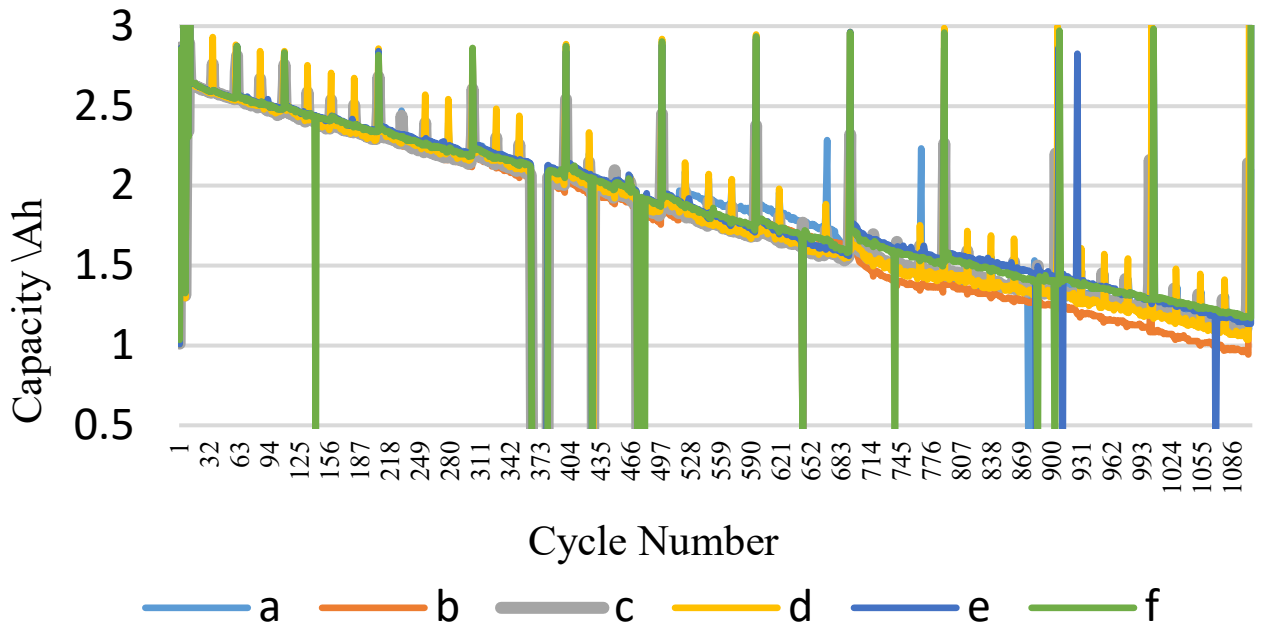

Figure 3. Detailed battery capacity curve of HNEI_18650_NMC_LCO_25C_0-100_0.5-1.5C.

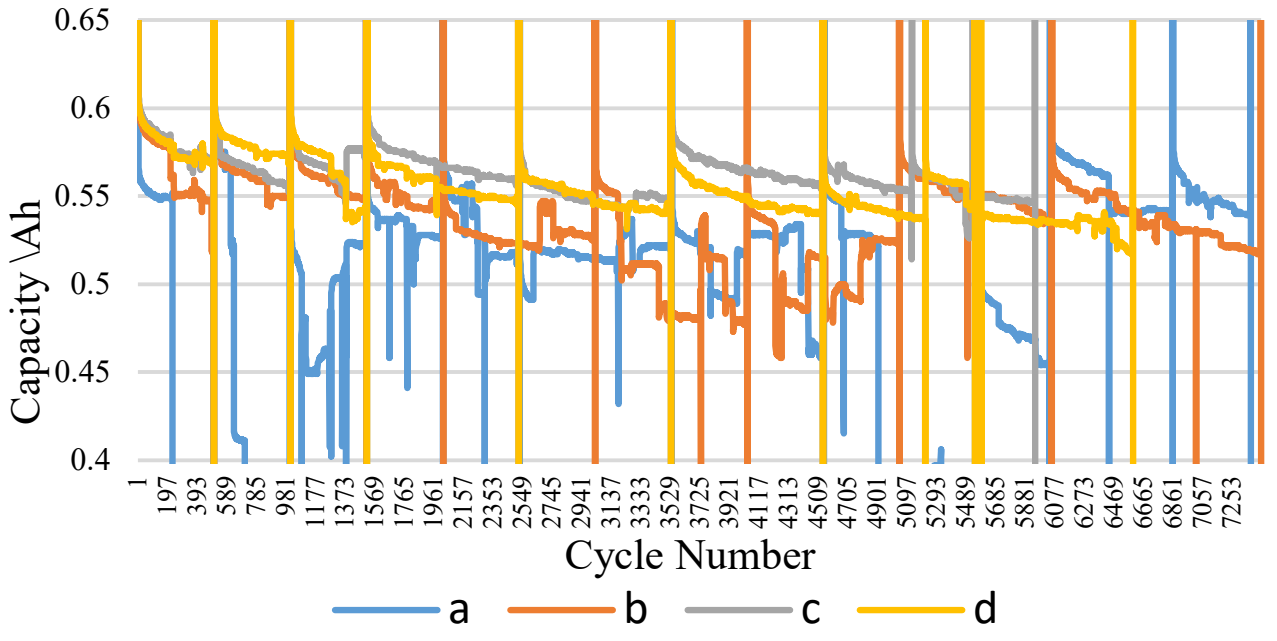

Figure 4. Battery capacity curve of SNL LFP 20-80 0.5C/0.5C 25C.

Fig. 5 presents the capacity curve of the same battery with Fig. 4 but the aging tests are under different discharge rates. The data applied here is the LFP battery at 35°C with 100% DOD. The charge rate is kept at 0.5C while the discharge rate varies by 0.5C, 1C, 1.5C and 2C for 4 battery aging tests respectively. From the figure, we can tell that the test with 3C discharge rate has the highest degradation among all other aging test while the degradation for 1C and 2C discharge rates is similar to each other. The 0.5C discharge rate results into the lowest degradation value. It seems that a higher discharge rate will lead to a higher degradation value.

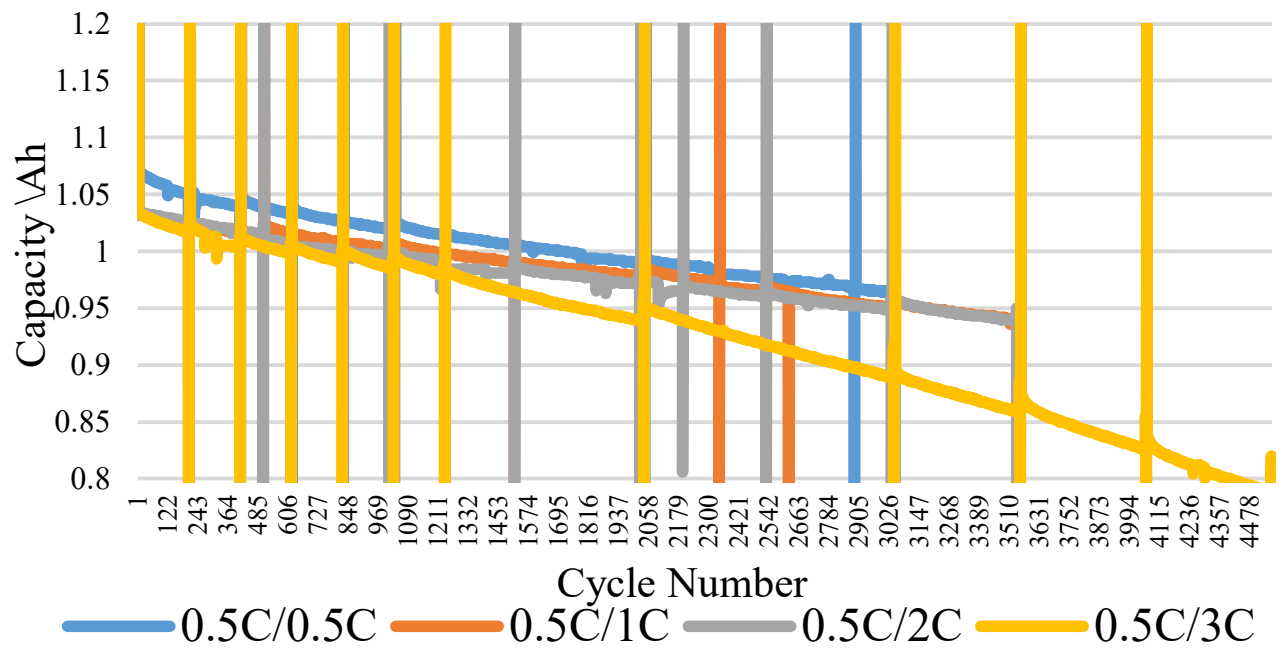

Figure 5. Battery capacity curve of SNL LFP 0-100 25°C.

There are also some battery degradation tests with different ambient temperature from the dataset. Fig. 6 shows the capacity curve under different operating temperature. All the aging tests applied here is under 1C charge rate and 100% DOD. From Fig. 6, we can observe that the higher ambient temperature leads to a higher battery degradation value. It takes more cycles for the battery test under 15°C to degrade to the same capacity level than the battery aging test under 25°C and 35°C. The lowest ambient temperature test that is available is 15°C. We believe that the extreme low temperature (lower than freeze point) will fast degrade the battery as well. Unfortunately, there is no such real degradation data to prove it. Also, if we increase the discharging rate to 2C, the previous conclusion from Fig. 6. is not true anymore. Fig. 7 shows that the battery has the lowest degradation when the ambient temperature is at 25°C. This may not be true if we switch to another type of battery.

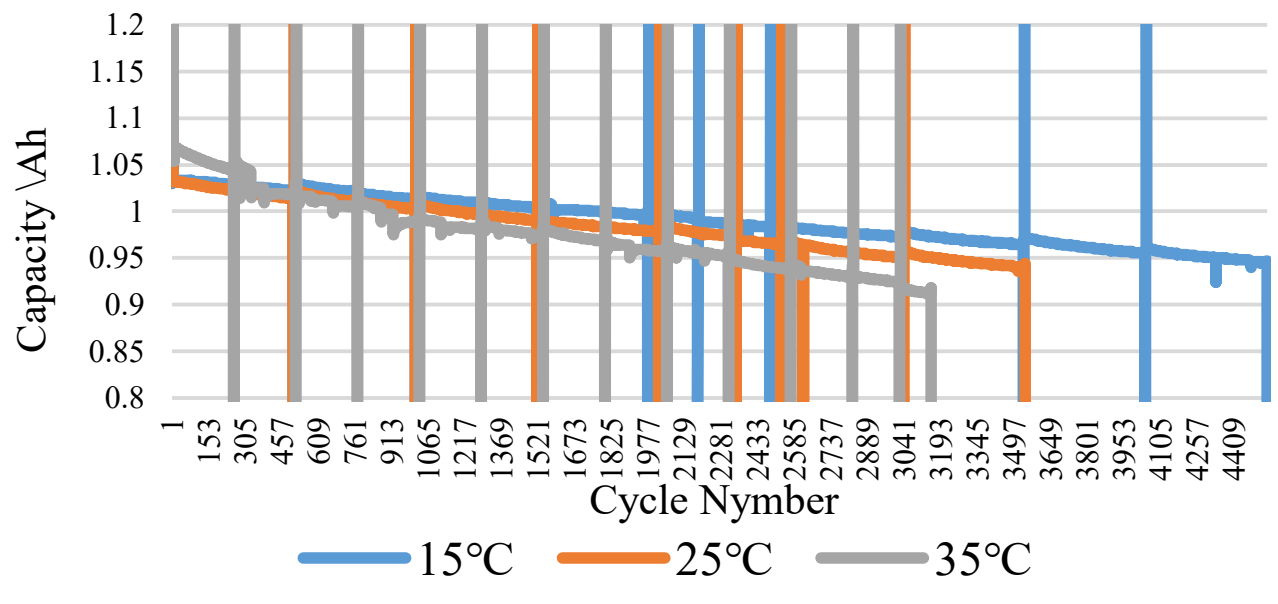

Figure 6. Battery capacity curve of SNL LFP 0-100 1C.

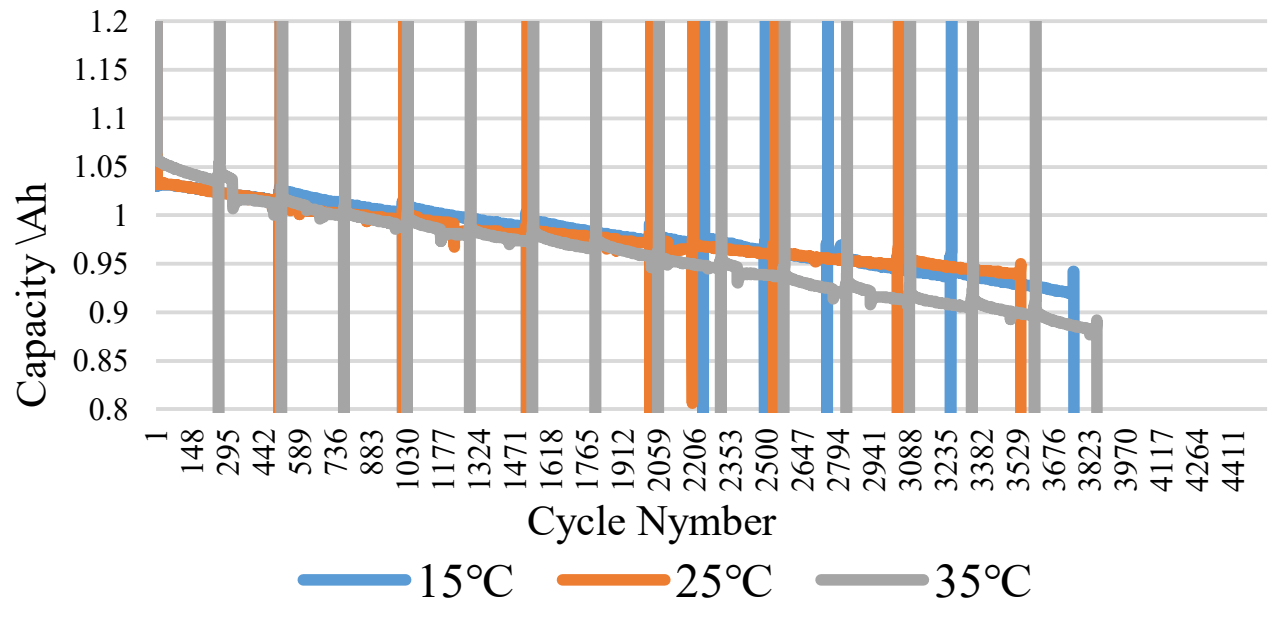

Figure 7. Battery capacity curve of SNL LFP 0-100 2C.

Fig. 8 and 9 are the dot plots of the degradation value. The negative value doesn't mean that there is a negative degradation, it is because the difference between the two continuous cycles is negative which is due to the stochastic characteristic of the lithium-ion battery. The calculated degradation values are positive for most of the cycles. After comparing Figs. 8 and 9, neglect the negative degradation numbers, we found that Fig. 9 has more positive degradation points, which indicates that the degradation is much faster at the 3C discharge rate. This matches the results in Fig. 5. It is worth noting that in the real battery degradation data, a negative battery degradation value is normal if we look at the two continuous cycles. Thus, a pre-processing is needed before applying the data.

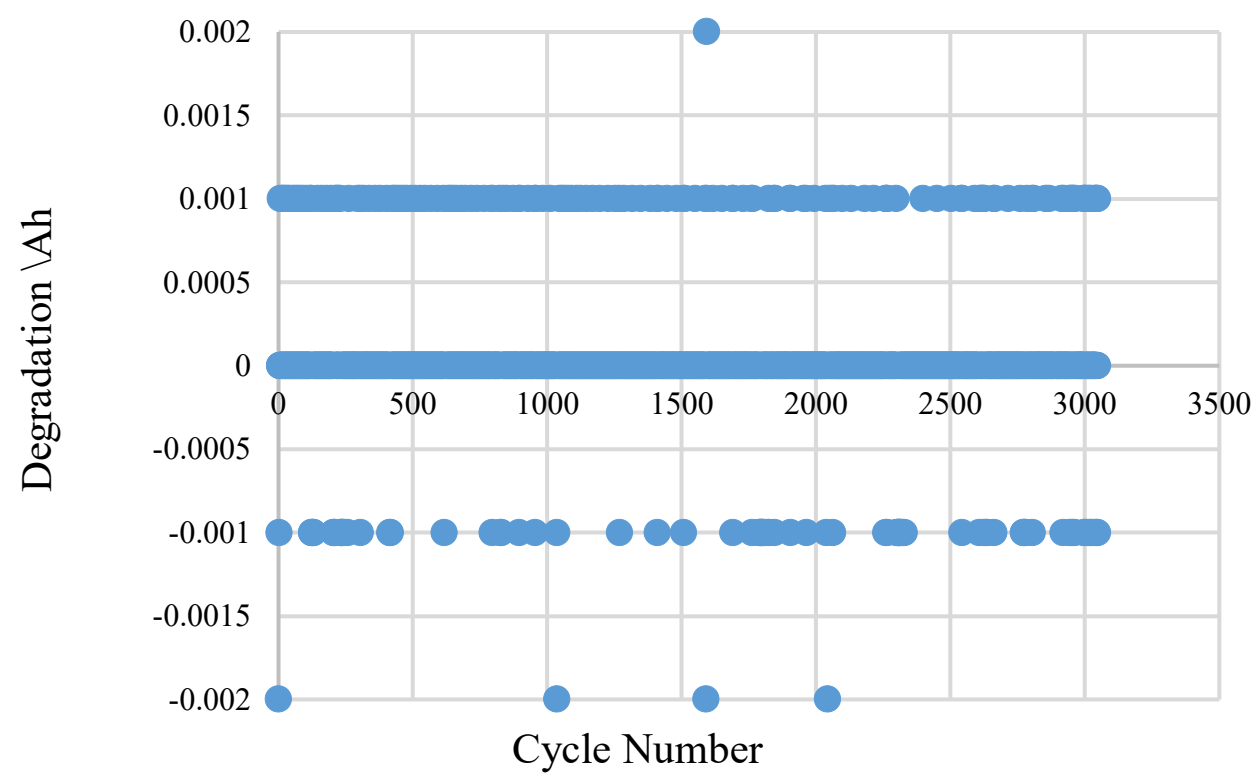

Figure 8. Battery degradation value of SNL LFP 0-100 0.5C 25°C.

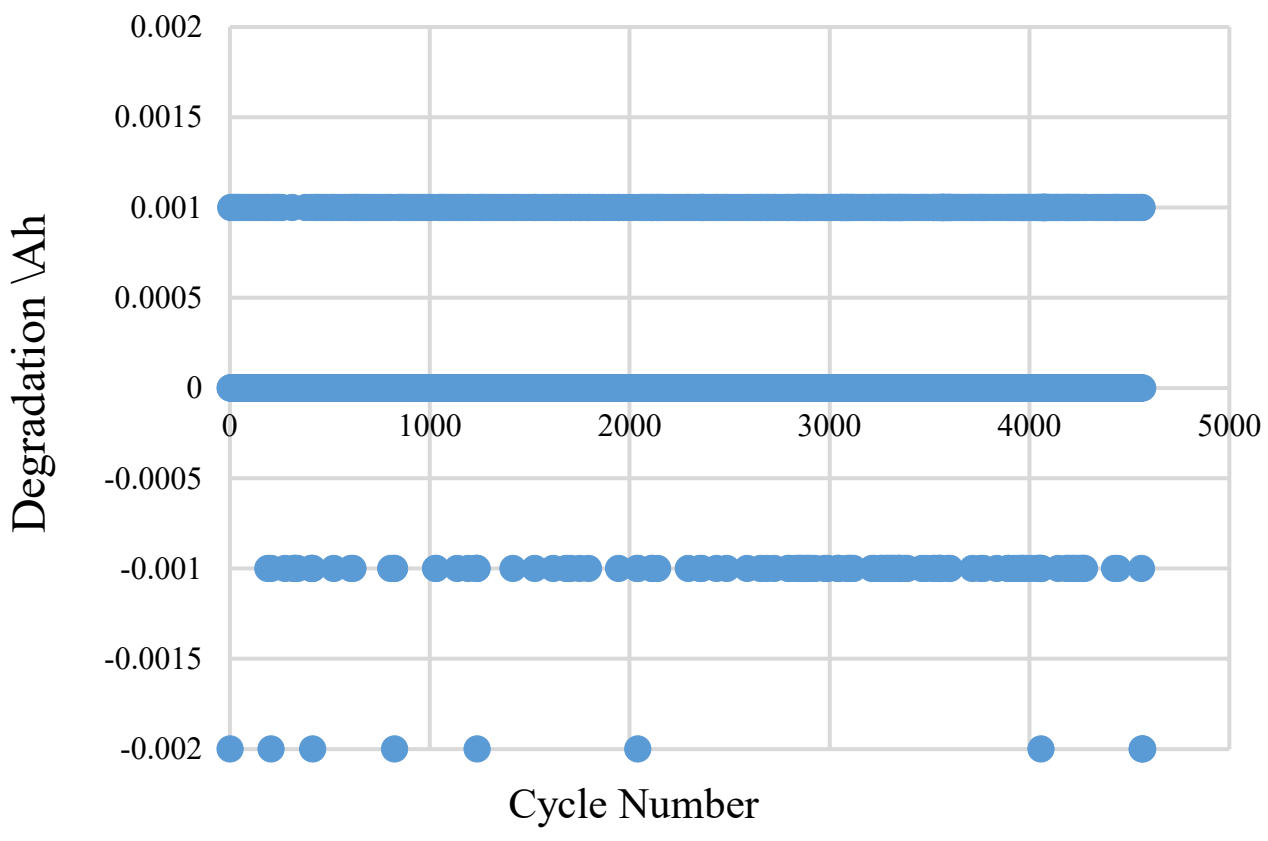

Figure 9. Battery degradation value of SNL LFP 0-100 3C 25°C.

## IV. Heuristic BDMs Comparison

The quality analysis of heuristic BDMs used in the literature is conducted in this section. There are three models studied in this paper: two heuristic models and one benchmark model. Model 1 and model 2 represent the two popular heuristic models mentioned in the previous section, referred to as linear BDM and DOD related BDM respectively. The benchmark model represents the averaged degradation value for each cycle from the real degradation data. The quality analysis is applied with different scenarios including different ambient temperature, charge or discharge rate and DOD. Note that the $c_{BD}$ and $c_{BD}^{DOD,t}$ for model 1 and model 2 are determined by one group of the aging tests from the real battery degradation data mentioned in section III. The battery degradation value of the benchmark model is the average degradation for each cycle based on the real battery degradation data under certain scenarios.

Fig. 10 presents the performance of the BDMs under different operating ambient temperature. The three groups of battery aging tests from the dataset that we applied here are the "100% 1C SNL 15 a", "100% 1C SNL 25 a" and "100% 1C SNL 35 a" respectively. We can observe that battery degradation predictions by model 1 are the same under different operating ambient temperature, so does model 2. This is because these heuristic popular BDMs do not consider the effectiveness of ambient temperature in their model. Compared to the benchmark model, the most accurate prediction on battery degradation value is model 2 at 15°C. This might be because the aging test that determines the $c_{BD}^{DOD,t}$ in model 2 is under a similar testing condition with the aging test of "100% 1C SNL 15". The benchmark model shows that the average degradation per cycle at 35°C is double more than 15°C. The degradation value at 25°C increases 25% from 15°C. It seems that a lower ambient temperature may lead to a lower battery degradation value. However, this may not be true; no further analysis can be conducted for now since there is no battery aging tests under extreme low ambient temperature in the battery archive dataset. From Fig. 10, we can conclude that the heuristic BDMs predict the battery degradation value with huge errors under different operating ambient temperatures. In other words, the heuristic BDMs are unable to perform well when dealing with varying ambient temperatures.

Fig. 11 shows the performance of BDMs under different discharge rates. The charge rate is fixed as 0.5C of the selected aging test data. "100% 1C SNL 15 b" and "100% 2C SNL 15 b" are the two groups of aging tests that are analyzed here. The DOD and the ambient temperature are fixed with 100% and 15°C. From Fig. 11, similar to the previous analysis, model 1 and model 2 perform the same on the battery degradation prediction with different discharge rates. Compared with the benchmark model, Model 2 predicts the same degradation value when the discharge rate is 1C. However, both model 1 and model 2 have low accuracies on the battery degradation prediction at the 2C degradation rate. Thus, we can conclude that the heuristic popular BDMs may not work well with different discharge rates. For benchmark model, the battery degradation value at 2C discharge rate is higher than it at 1C. We believe the 3C discharge rate will lead to a much higher degradation value. However, the database doesn't have enough 3C discharge rate data to be analyzed.

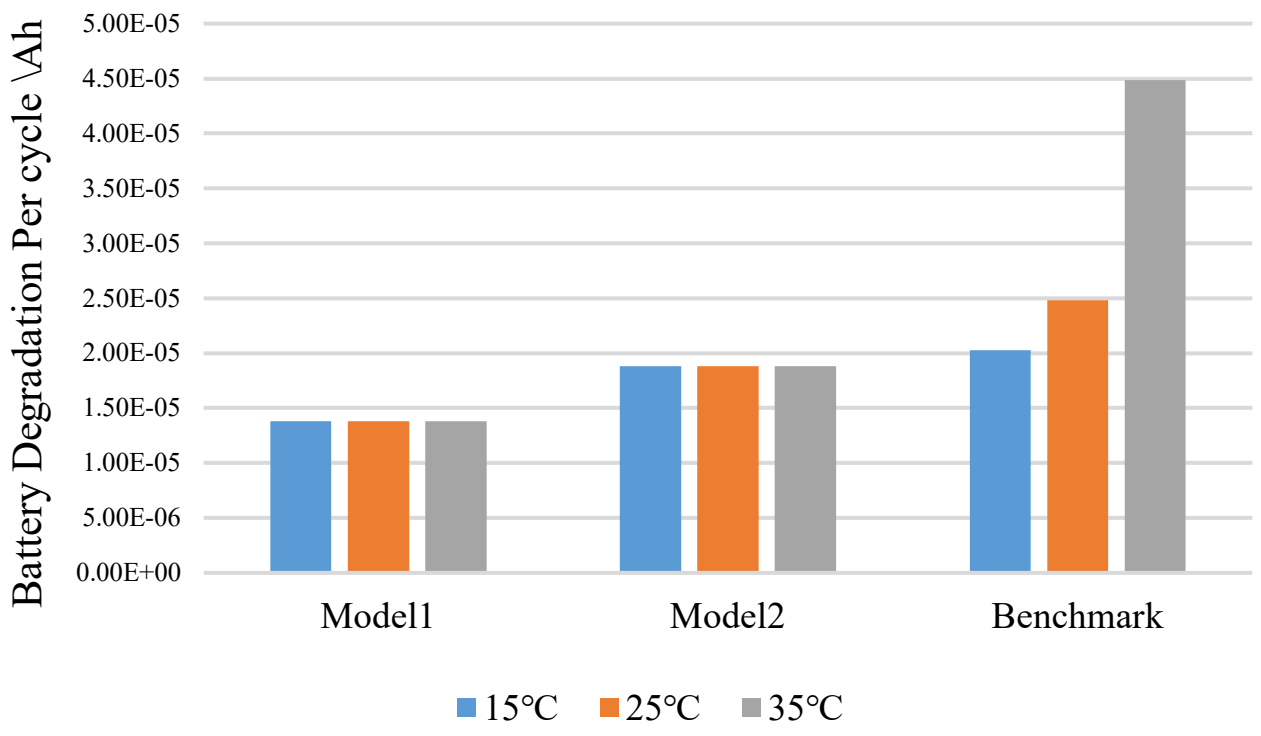

Figure 10. Degradation comparison under different ambient temperatures.

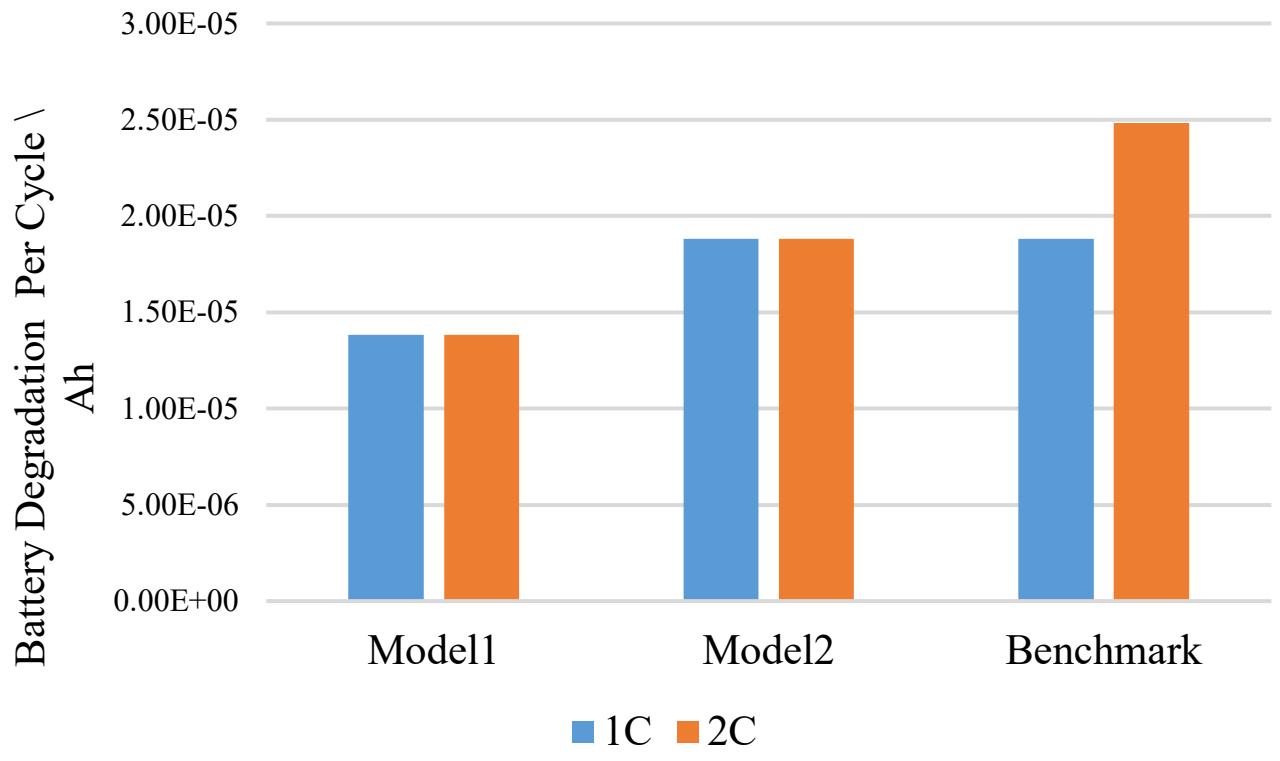

Figure 11. Degradation comparison under different discharge rates.

The DOD is also a key factor that affects the battery degradation value. Fig. 12 presents the results of the BDMs with different DOD value. Three battery aging tests with different DOD values of 100%, 60% and 40% respectively from the battery degradation dataset are used here for the analysis. It is clear that all three models share the same pattern that higher DOD values result into higher degradation values. This results also meet the expectation of the heuristic BDMs. However, the accuracy of the prediction of model 1 and model 2 is low. The scenarios that determine $c_{BD}$ and $c_{BD}^{DOD,t}$ are different from the aging tests that are selected here, which leads to the inaccurate battery degradation prediction. In other words, the aging test that applied to determine the degradation parameter in the heuristic models is the key to improve the prediction accuracy. It may need to update the $c_{BD}$ and $c_{BD}^{DOD,t}$ each time to keep the prediction accurate. However, the open source battery degradation data are limited and such data are not always available practically. Thus, a better BDM is required to accurately predict the battery degradation with limited degradation data resource.

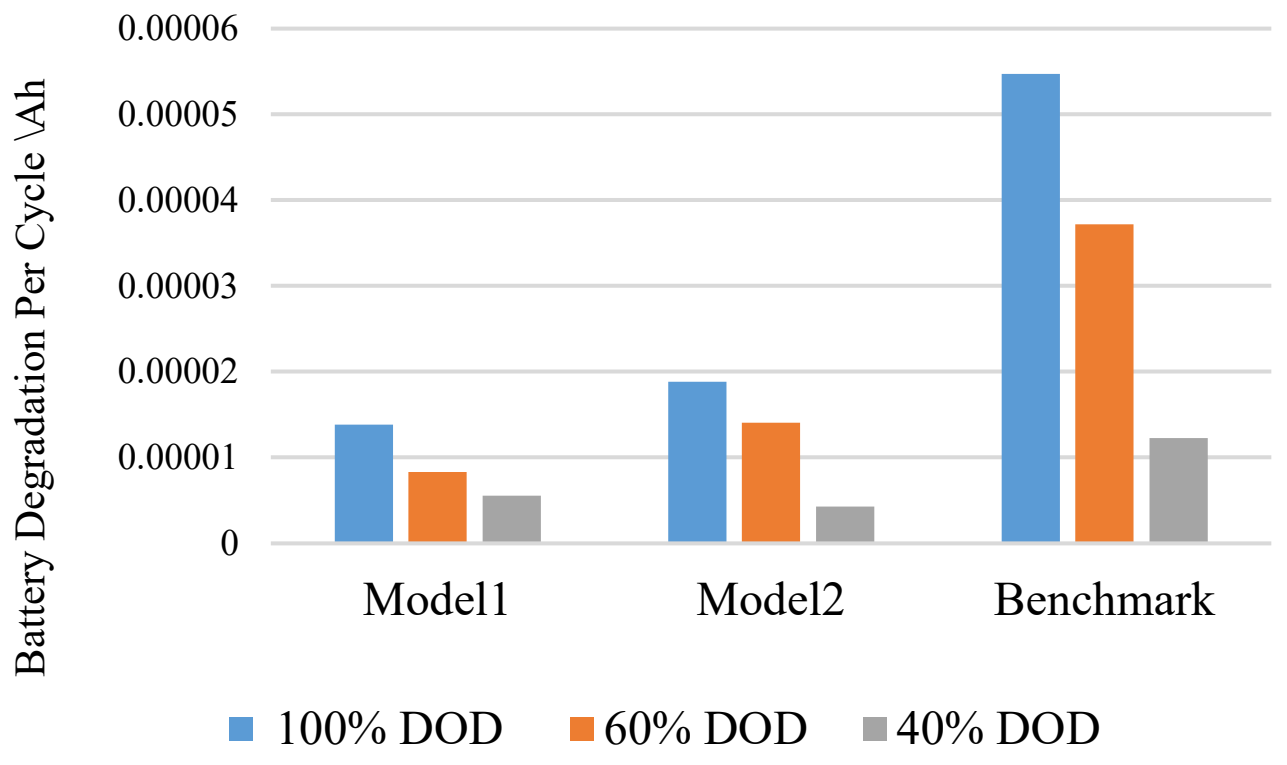

Figure 12. Degradation comparison under different DOD values.

## V. Conclusion

A quality analysis is conducted on the popular battery degradation models in this paper. The results show that the popular heuristic BDMs often quantify the battery degradation very incorrectly under different scenarios, which may lead to a substantial unnecessary reduction of battery lifetime. Thus, a comprehensive BDM that considers all the battery aging factors is needed to accurately predict the battery degradation value for use of future energy management systems. The chemical characteristics make battery degradation hard to predict. However, the analysis of the real battery degradation data shows that lower charge or discharge rate and smaller DOD will lead to lower degradation values. Also, we can conclude that the battery performs differently under different ambient temperatures or different battery types. To summarize, this paper provides an overview of the performance of popular heuristic BDMs and conducts the real battery degradation analysis.

## Acknowledgment

This work is sponsored by the Grants to Enhance and Advance Research (GEAR) program by Division of Research at the University of Houston.